# CLUSTER ABUNDANCE FROM PERTURBATION SPECTRA WITH BROKEN SCALE INVARIANCE

VOLKER MÜLLER

*Astrophysical Institute Potsdam*
*An der Sternwarte 16, D-14482 Potsdam*
*vmueller@aip.de*

## ABSTRACT

A double inflationary model with enhanced power at large scales — Broken Scale Invariant perturbations (BSI) — leads to a promising scenario for the formation of large-scale structure. We use high-resolution PM simulations with a model for the thermodynamic evolution of baryons. Galactic halos are identified as peaks in the distribution of mass points which had time to cool providing a physical biasing mechanism. The clustering properties of galactic halos are in agreement with both small-scale and large-scale characteristics of the galaxy distribution, as variances of counts in cells, 'galaxy' correlation functions, 'cluster' abundances and 'cluster-cluster' correlation functions, and velocity dispersions at small scales.

## 1. Introduction

During the last years, increasing evidence has been accumulated that the standard CDM model in a $\Omega = 1$ universe as suggested by inflation leads to inconsistencies. First, the APM survey[1] showed more power on scales larger than the galaxy correlation length. Later, the COBE measurements[2] of the microwave background anisotropy provided evidence, that the CDM model would require an antibiasing for conformity with the observed galaxy clustering. Later, a large number of observations came in difficulties with the CDM model. Therefore, additional admixtures of hot besides cold dark matter, or a cosmological term, or an open cosmological model have been proposed. We took up the old idea of a multi-inflationary scenario[3]. The model decouples the structure formation on large and small scale, producing a power spectrum with a break at a characteristic scale[4]. In addition to the amplitude, it requires two parameters: first, the break height being proportional to the ratio of the effective masses during the inflationary evolution; second, the epoch of the transition between the inflationary phases which is related to the initial energy density of the inflaton field. This epoch determines the scale length $l_s \equiv 2\pi/k_s$ of the break.

We consider a special double inflation spectrum, which — using the linear theory — provides a good description to the large scale matter distribution in the universe[5], also the APM angular correlation function can be well fitted[6]. A description of the primordial potential spectrum is given by the approximate formula (it does not describe the typical oscillations generated during the intermediate power law stage):

$$k^3 P_\Phi(k) = \begin{cases} 4.2 \ 10^{-6} [\log(k_s/k)]^{0.6} + 4.7 \ 10^{-6} & \text{for} \quad k < k_s \\ 9.4 \ 10^{-8} \log(k_f/k) & \text{for} \quad k > k_s \end{cases} \quad (1)$$

with $k_s = (2\pi/24)h^{-1}\text{Mpc}$, $k_f = e^{56}h^{-1}\text{Mpc}$. We fold this spectrum with the standard CDM transfer functions (using a dimensionless Hubble constant $h = 0.5$) to get the power spectrum shown in Fig. 1. In the left we compare this BSI spectrum (solid line) with the standard CDM model (dashed line), a $\Lambda$CDM ($\lambda \equiv \Lambda/3H^2 = 0.8$, $\Omega = 0.2$, dotted line) and a MDM model ($\Omega_{\text{CDM}} = 0.7$, $\Omega_\nu = 0.2$, $\Omega_b = 0.1$, dash-dotted line). The standard CDM spectrum has more power at the scales relevant for galaxy formation, while BSI mostly resembles MDM at large scales, but has more power on small scales, almost as high as $\Lambda$CDM.

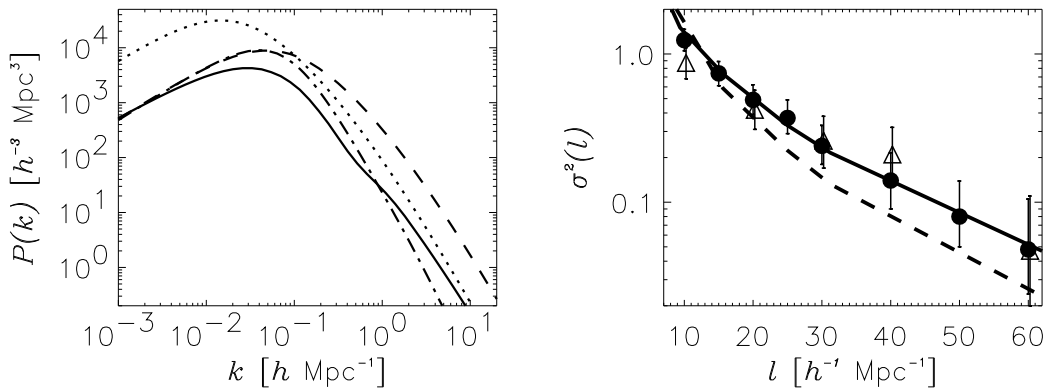

Figure 1: Power spectra (left) and variances of 'galaxy' count in cells.

We have simulated the formation of large–scale structure employing a particle-mesh code[7] and using a resolution of $512^3$ cells and $256^3$ particles for BSI[8], comparisons are made for COBE-normalised CDM with $256^3$ cells and $128^3$ particles. The box sizes lie between 25 $h^{-1}$Mpc and 500 $h^{-1}$Mpc to get a wide dynamical range (cp., Gottlöber, this volume). On the right of Fig. 1 we compare simulated count-in-cell variances at $z = 0$ as function of radius with observations in the IRAS catalogue[9] (triangles) and the Stromlo-APM survey[10] (circles). There the BSI curve (solid line) has been shifted vertically by a "biasing" factor $b = 2$ in order to normalize fluctuations to observations at $8h^{-1}$Mpc, while the CDM curve (dashed line) has $b = 0.9$. On scales $(15 - 50)h^{-1}$Mpc, the CDM fluctuations are too low (outside of the 2-$\sigma$ error range), at larger scales increase the errors. In contrast, the BSI simulations fit the data quite well, thereby testing the slope of the proposed spectrum.

The used code includes a model for thermodynamic evolution of baryons (heating by shocks during the first anisotropic collapse, the standard cooling of the primordial matter, and a reheating of cooled particles to $5 \times 10^4$ K with a fixed propability of 0.85 of the baryons, say by supernovae, cp. Kates, this volume). The thermodynamic information is attached to the dark matter, not allowing a description of genuine hydrodynamic effects. We get an estimate of the local gas temperature used to describe the condensation of baryons which helps to identify different classes of objects in the universe.

## 2. Galaxy halos

After the first stage of pancake formation and growth, the particles virialize in potential wells possibly giving rise to galaxies. The galaxy finding procedure used is a modified density peak prescription. Based on previous experience[7], we search for local maxima in the density contrast of cold particles exceeding a threshold $\delta_{\text{th}}$. We expect that shallower maxima do not lead to galaxies. The mass of 'galaxies' is determined by summing the contributions of all particles within a characteristic radius. For each simulation, several halo catalogues were constructed using thresholds $\delta_{\text{th}} = (1.5 - 3)\sigma_\delta$.

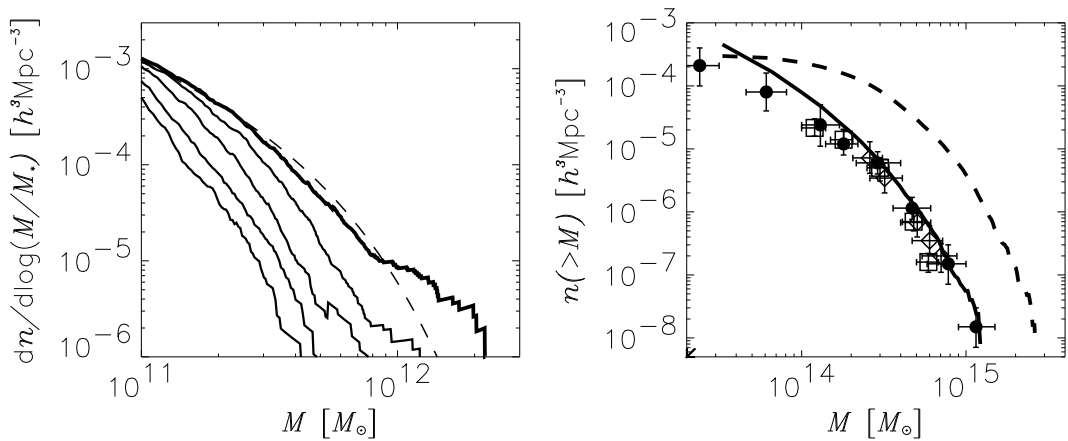

Figure 2: 'Galaxy' mass spectra (left) and 'cluster' mass function (right).

In Fig. 2 we show 'galaxy' mass functions obtained in a $25h^{-1}$Mpc simulation at redshifts $z = 2, 1.5, 1, .5, 0$ (curves from below). The dashed interpolation at $z = 0$ corresponds to a Schechter fit

$$\Phi(M)\mathrm{d}M = \Phi_*(M/M_*)^{-p}\exp(-M/M_*\mathrm{d}M, \qquad (2)$$

with $M_* = 3 \times 10^{11} M_\odot$, $\Phi_* = 10^{-3} h^3 \text{Mpc}^{-3}$ and $p = 1$. These parameters are comparable to the observations[10], but there, the mass spectra are steeper (the simulated CDM galaxy mass spectra of the same resolution are still much flatter, providing too high numbers of heavy galaxies). This result is partly due to overmerging typical for our dissipationless DM simulations and the PM dynamics. However, the strong overabundance of high mass halos is typical for the large power in the COBE-normalised CDM spectrum. We do not intend to make more detailed comparison with observed mass spectra of galaxies, but we use the identified halos with reasonable number density to compare their clustering properties with the observed large scale structure of the universe.

In the left of Fig. 3, we show the two-point correlation function for the galaxy catalogue constructed from a 200 $h^{-1}$Mpc simulation (dash-dotted line). It is well described by a power law $\xi = (r/r_0)^{-\gamma}$ for $1h^{-1}$Mpc $< r < 15h^{-1}$Mpc , with slope $\gamma \approx 1.6$ and correlation length $r_0 = 5.5h^{-1}$Mpc (straight line), in agreement with the CfA-catalogue[11]. The observed slope is influenced by peculiar velocity distortions as demonstrated by the dashed curve, extracted from all DM particles and imposing a linear bias $b = 1.5$. This is approximately the bias of a typical galaxy catalogue. The upper solid line is the correlation function of 'cluster' halos selected from the $500h^{-1}$Mpc simulation, which is enhanced by the same biasing factor; the slope and the correlation length $r_0 \approx 15h^{-1}$Mpc are typical for cluster data (cp. Bahcall, and Böhringer, this volume). As in the case of the mass spectra, in the COBE-normalized CDM simulations the correlation function has a too steep slope (this is characteristic for low bias CDM[12]).

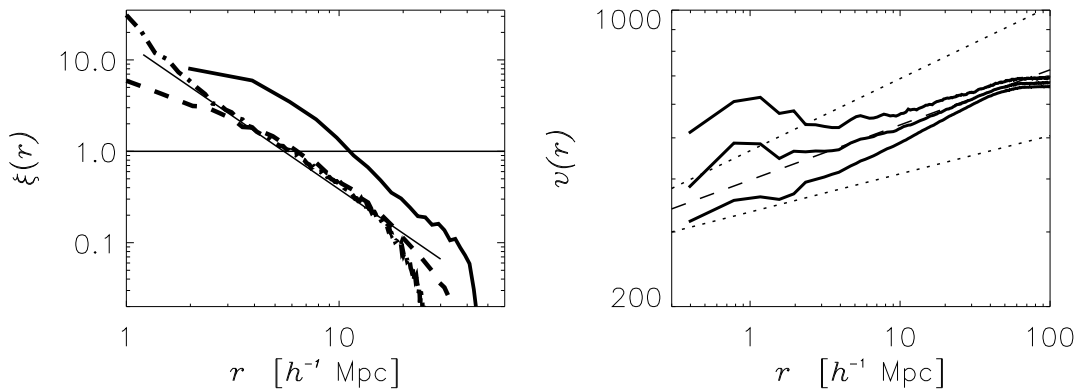

Figure 3: Correlation functions (left) and relative peculiar velocities (right).

## 3. Analysis of Galaxy Clusters

Clusters of galaxies provide a sensitive test of cosmological scenarios. Their formation is connected with the recent decoupling of large masses from the cosmic evolution. X-ray surveys[13] and optical cluster catalogues are used to define a general 'cluster' mass function[14]. It tests the slope of the power spectrum at the break scale. Based on the Press–Schechter theory and using BSI primordial power spectra, we derived strong bounds on the break scale $k_s$[15]. To verify this conclusion we identify galaxy clusters in the $500h^{-1}$Mpc simulations with a search radius of 1.5 $h^{-1}$Mpc ; it gives a reasonable cluster correlation radius and a positive correlation range up to 80 $h^{-1}$Mpc (cp. Fig. 3), like in observations. In Fig. 2 (right), we compare the cluster mass function with the data[14], both the slope and the normalization fit the data (the solid line denotes BSI), only at $M < 6 \times 10^{13} M_\odot$ one notices a slight overproduction of cluster halos. Our estimate does not use any additional parameter besides the input perturbation spectrum, in particular it does not depend on the details of the thermodynamic assumptions in the simulations (we collect all particles in the cluster halos). The cluster formation is a quite recent process for the BSI spectra. In the contrary, the CDM model leads to a strong overproduction of halos (cp. the dashed line), the mass function is to steep at cluster scales, and clusters are formed at high redshifts, contrary to observations.

The velocity dispersion of 'galaxies' at different scales is an important test of the form of the power spectrum. We show the relative peculiar velocities as function of the distance on the right of Fig. 3. The dashed and dotted lines are the mean and 1-$\sigma$ deviations of the velocity dispersion as determined by the anisotropy of the small scale velocity dispersion[11]. The three simulation curves correspond to 'galaxy' catalogues with different thresholds corresponding to mass limits $M > (5, 8, 11) \, 10^{11} M_\odot$ (from below). The different velocity dispersions are due to the higher probability of finding large mass halos in dense clusters, both due to merging and over-merging. As in the case of the correlation and mass functions, the studied CDM model leads to much higher small scale velocity dispersions[8]. The simulated velocity seggregation as shown in Fig. 3 represents a sensible test of the model.

The primordial BSI power spectrum provides definitive predicions for the galaxy clustering. Our 'galaxy' selection procedure should be compared with full hydrodynamical simulations to test its trustworthiness. Certain statistics of large-scale structure as the correlation functions and the velocity fields are insensible to this scheme. They show that the BSI model passes many cosmological tests. The most important question as compared with alternatives as CHDM or $\Lambda$CDM is the redshift dependence of clustering, and in particular, a quantitative comparison with predictions of the development of the halo mass function as discussed in Fig. 2.